\documentstyle[aps,multicol,epsf]{revtex}
\tighten
\begin{document}
\draft
\title{Smearing of Coulomb Blockade by Resonant Tunneling}
\author{Thomas Gramespacher and K.A.~Matveev}
\address{Department of Physics, Duke University, Durham, NC 27708-0305}
\date{\today}
\maketitle
\begin{abstract}
  We study the Coulomb blockade in a grain coupled to a lead via a
  resonant impurity level. We show that the strong energy dependence of
  the transmission coefficient through the impurity level can have a
  dramatic effect on the quantization of the grain charge.  In particular,
  if the resonance is sufficiently narrow, the Coulomb staircase shows
  very sharp steps even if the transmission through the impurity at the
  Fermi energy is perfect. This is in contrast to the naive expectation
  that perfect transmission should completely smear charging effects.
\end{abstract}
\pacs{PACS numbers: 73.23.Hk, 73.40.Gk, 73.23.-b}

\begin{multicols}{2}

The charge on an isolated metallic grain is quantized in units of the
electron charge $e$.  Even if the grain is weakly coupled to an electrode,
so that electrons can occasionally hop from the electrode to the grain and
back, the charge on the grain still remains to a large extend quantized.
This phenomenon, known as Coulomb blockade, has in recent years been
widely investigated, both theoretically and
experimentally\cite{averin91,grabert92,kouwenhoven97}. One quantity of
interest is the average charge on the grain as a function of the voltage
applied to a nearby gate.  For a very weakly coupled grain this function
shows very sharp steps, the so called Coulomb staircase. As has been shown
in recent experiments\cite{lafarge93,berman99}, the charge on such a grain
can directly be measured using a single electron transistor that is
capacitively coupled to the grain. The reason for the charge on the grain
to be quantized is that it costs a finite energy $E_C=e^2/2C$ to charge
the capacitance $C$ formed by the grain and its environment.  Charge
quantization effects therefore become visible as soon as the temperature
$T$ is lowered below $E_C$.  From now on we assume that the temperature is
zero.

As the coupling of the grain to the lead is made stronger and stronger,
the sharp steps of the Coulomb staircase are more and more smeared out.
One usually assumes that all features of charge quantization completely
disappear as soon as the coupling of the grain to the lead is via a
perfectly transmissive channel\cite{vandervaart93,matveev95,nazarov99}.
However, this is only the case if the transmission probability from lead
to grain is unity in an energy interval much broader than the charging
energy $E_C$ around the Fermi energy. We will show below that perfect
transmission in a narrow energy interval is not sufficient to effectively
smear out charging effects. We study a model where perfect transmission is
achieved using a resonant impurity level connecting the grain to the lead.
The transmission probability through such a resonant impurity level is
strongly energy dependent and is characterized by the width $\Gamma$ of
the resonance.  Embedding the level between sufficiently high tunneling
barriers can give a small $\Gamma\ll E_C$. In this regime, one can achieve
perfect transmission between the grain and the lead at the Fermi level and
still have a nearly perfectly sharp Coulomb staircase.  As the
resonance\hspace{.8mm} is\hspace{.8mm} made\hspace{.8mm} wider,\hspace{1mm}
the\hspace{.8mm} sharp\hspace{.8mm} steps\hspace{.8mm} of\hspace{.8mm}
the\hspace{.8mm} Coulomb
\begin{figure}
\epsfxsize0.8\columnwidth
\epsffile{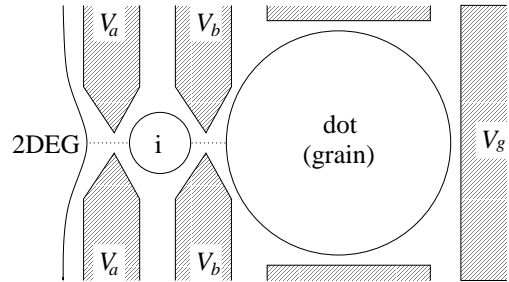}
\vspace{4ex}
\caption{A possible experimental setup using a GaAs heterostructure. The
  small quantum dot plays the role of the impurity level, and the larger
  dot represents the grain. The charge $Q$ of the large dot is controlled
  by the gate voltage $V_g$.  The voltages $V_a$ and $V_b$ control the
  heights of the barriers between the dots and the lead and thus the width
  of the resonance.
\label{schema}}
\end{figure}
\noindent
staircase start to be
smeared out and will eventually disappear at $\Gamma\gg E_C$.

The experimental setup we have in mind could, e.g., be a metallic grain,
separated from a massive electrode by a thin insulating layer containing
resonant impurity states near the Fermi level. Another possibility would
be a double quantum dot system, where one of the dots is considerably
smaller than the other, Fig.~\ref{schema}.  The level spacing in the small
dot then exceeds by far the charging energy of the larger one, so that the
small dot can only be occupied by zero or one electron.  The small dot
plays then the role of the impurity. The advantage of this setup is that
by tuning the gate voltages the barrier heights between the two dots and
the lead can be adjusted.  In addition, the effective energy of the
impurity can be shifted by a gate which couples only to the small dot.

The model we consider is described by the following Hamiltonian,
\begin{equation}
H=H_0+H_{li}+H_{ig}. \label{genham}
\end{equation}
Here the Hamiltonian $H_0$ describes the lead, the impurity and the
grain,
\begin{eqnarray}
H_0 & = & \sum_{k\sigma}\epsilon_k a_{k\sigma}^\dagger a_{k\sigma}
+\sum_{\sigma}\epsilon a_\sigma^\dagger a_\sigma \nonumber \\
 & & +\sum_{p\sigma}\epsilon_p a_{p\sigma}^\dagger
 a_{p\sigma}+E_C(\hat{n}-N)^2, 
\end{eqnarray}
and the coupling of the impurity to the other two electrodes is described
by the tunneling Hamiltonians:
\begin{eqnarray}
H_{li}&=&\sum_{k\sigma} (t_k a^\dagger_{k\sigma} a_\sigma + h.c.), \\
H_{ig}&=&\sum_{p\sigma} (t_p a^\dagger_{p\sigma} a_\sigma + h.c.).
\end{eqnarray}
Here, $a_{k\sigma}$, $a_\sigma$, and $a_{p\sigma}$ are the annihilation
operators for electrons of spin $\sigma$ in the lead, impurity, and the
grain, respectively.  The operator $\hat{n}=\sum(a^\dagger_{p\sigma}
a_{p\sigma}-\langle a^\dagger_{p\sigma}a_{p\sigma}\rangle_0)$ counts the
number of electrons on the grain relative to its expectation value for the
uncoupled system.  The parameter $N$ is proportional to the gate voltage
$V_g$, namely $N=C_g V_g/e$, where $C_g$ is the capacitance between the
grain and the gate electrode.  In this model we neglected the interaction
of electrons on the impurity site. We will later include a strong Coulomb
repulsion on the impurity.

In the absence of tunneling, $H_{li}=H_{ig}=0$, the charge $\langle
Q\rangle$ on the grain is a multiple of the electron charge $e$, and
thus is perfectly quantized.  We will now investigate how the coupling via
the impurity level affects the charge quantization on the grain. We assume
that the coupling of the impurity to the grain is weak, so that it is
sufficient to treat $H_{ig}$ in perturbation theory.  We can then take
advantage of the fact that the system of an impurity coupled to a lead is
non-interacting, and can therefore be easily solved exactly. Let
$|0\rangle$ denote the ground state of the Hamiltonian (\ref{genham}) with
$H_{ig}=0$.  The first order correction $|\delta\psi\rangle $ to
$|0\rangle $ is
\begin{equation}
|\delta\psi\rangle=-i\int_{-\infty}^0 dt H_{ig}(t)|0\rangle\, ,
\end{equation}
where $H_{ig}(t)$ is the time evolution of the coupling taken in the
interaction representation.  The expectation value of the charge on the
dot is then to second order in the coupling to the impurity
\begin{eqnarray}
\langle Q\rangle
& = & -e\sum_{p\sigma}|t_p|^2 \int_{-\infty}^0 dt_1\int_{-\infty}^0
dt_2 \nonumber \\
& & \times \left[\langle
a_\sigma(t_2)a^\dagger_\sigma(t_1)\rangle
 \langle a^\dagger_{p\sigma}(t_2)a_{p\sigma}(t_1)
\rangle\right.
\nonumber \\
& & -\left.\langle a_\sigma^\dagger(t_2)a_\sigma(t_1)\rangle\langle
a_{p\sigma}(t_2)a^\dagger_{p\sigma}(t_1)\rangle \right] , 
\label{dqsecondord}
\end{eqnarray}
where the averages are taken over the ground state of the uncoupled system.
The Green's functions of the isolated grain can be calculated easily:
\begin{eqnarray}
\langle a^\dagger_{p\sigma}(t_2)a_{p\sigma}(t_1)\rangle
& = & \theta(-\epsilon_p)e^{i(\epsilon_p-U_{-1})(t_2-t_1)} , \nonumber \\
\langle a_{p\sigma}(t_2)a^\dagger_{p\sigma}(t_1)\rangle
& = & \theta(\epsilon_p) e^{-i(\epsilon_p+U_1)(t_2-t_1)} . \nonumber
\end{eqnarray}
The Green's functions of the non-interacting impurity/lead system can be found
by solving their equations of motion:
\begin{eqnarray}
\langle
a_\sigma(t_2)a^\dagger_\sigma(t_1)\rangle & = &
\int_0^\infty\frac{d\omega}{\pi}\frac{\Gamma_l e^{-i\omega (t_2-t_1)}} 
{(\omega-\epsilon)^2+\Gamma_l^2}, \nonumber \\ \langle
a^\dagger_\sigma(t_2)a_\sigma(t_1)\rangle    & = &
\int_0^\infty\frac{d\omega}{\pi} \frac{\Gamma_l e^{-i\omega (t_2-t_1)}}
{(\omega+\epsilon)^2+\Gamma_l^2}. \nonumber
\end{eqnarray}
Using these relations, the integrals in Eq.~(\ref{dqsecondord}) can be
evaluated and yield
\end{multicols}
\begin{eqnarray}
\langle Q\rangle & = &
2e\frac{\Gamma_g}{\pi^2}\left\{\frac{U_1-\epsilon}
         {\Gamma_l^2+(U_1-\epsilon)^2}\left[
\frac{\pi}{2}-\arctan\left(\frac{\epsilon}{\Gamma_l}\right)
           +\frac{\Gamma_l}{U_1-\epsilon}
\ln\frac{\sqrt{\epsilon^2+\Gamma_l^2}}{U_1}\right]\right. \nonumber \\
& & \left.-\frac{U_{-1}+\epsilon}{\Gamma_l^2+(U_{-1}+\epsilon)^2}
  \left[\frac{\pi}{2}
+\arctan\left(\frac{\epsilon}{\Gamma_l}\right)
         +\frac{\Gamma_l}{U_{-1}+\epsilon}
\ln\frac{\sqrt{\epsilon^2+\Gamma_l^2}}{U_{-1}}\right]\right\} . 
 \label{dqmostgen}
\end{eqnarray}
\begin{multicols}{2}
In this result we have five independent energy scales, namely the couplings
$\Gamma_g=\pi\sum_p|t_p|^2\delta(\epsilon_p)$ and
$\Gamma_l=\pi\sum_k|t_k|^2\delta(\epsilon_k)$, the energy $\epsilon$ of
the impurity and, finally, the Coulomb energies $U_{-1}=E_C(1+2N)$ and
$U_1=E_C(1-2N)$, which have to be paid if an electron is removed from or
added to the grain, respectively.
We assume the energy spectrum in the lead as well as in the grain to be
continuous. This implies that the grain is sufficiently large, so that the
level spacing $\Delta$ is much smaller than all other relevant energy scales,
$\Delta\ll\Gamma_{l,g},U_{\pm 1}$. In this regime, the
mesoscopic fluctuations of the coupling elements $t_p$ will naturally
average out in the expression for $\Gamma_g$.

Let us now discuss this result.
Clearly, the charge smearing is linear in $\Gamma_g$, since we treated the
coupling $H_{ig}$ only to second order.  The coupling $H_{li}$ has been
accounted for to all orders. Note, that even if $\Gamma_l=0$, i.e., the
lead is decoupled from the rest of the system, the charge smearing does
not vanish,
\begin{equation}
\langle Q\rangle=\mp 2e\frac{\Gamma_g}{\pi}\frac{1}{U_{\mp 1}\pm \epsilon}, 
\label{dqonlyimp}
\end{equation}
where the top and bottom signs correspond to positive and negative
$\epsilon$, respectively.  This result can of course be easily obtained by
performing a second-order perturbation theory with respect to $H_{ig}$ in
a system decoupled from the lead, $H_{li}=0$.  Clearly, the processes of
multiple tunneling between the impurity and the grain result in
corrections which are small in the parameter $\Gamma_g/U_{\mp 1}$.  Therefore
the lowest-order perturbation theory in $H_{ig}$ employed in the
derivation of Eqs.~(\ref{dqonlyimp}) and (\ref{dqmostgen}) is applicable
away from the degeneracy points, i.e., at $\Gamma_g\ll U_{\pm1}$. 

As a next step, we investigate how the charge smearing is affected by the
coupling to the lead, assuming now that both couplings, $\Gamma_l$ and
$\Gamma_g$, are finite. Let us first consider the case where the impurity
level is far above the Fermi surface, i.e., we assume that
$\epsilon\gg\Gamma_{l,g},U_{\pm 1}$. Then Eq.\ (\ref{dqmostgen}) can be
simplified to
\begin{equation}
\langle Q\rangle=-2e\frac{\Gamma_g}{\pi}\frac{1}{\epsilon}
                 +2e\frac{\Gamma_g\Gamma_l}
{\pi^2\epsilon^2}\ln\frac{U_{-1}}{U_1}. 
\label{dqpertlargeeps}
\end{equation}
The first term on the right hand side coincides with Eq.~(\ref{dqonlyimp})
in the limit of large $\epsilon$ and is only due to the escape of
electrons from the grain to the impurity. The second term is due to
transfer of electrons from the grain to the lead and is equivalent to the
lowest-order result for charge smearing for a grain coupled to a lead via
a tunneling barrier \cite{Glazman90}.  Comparing
Eq.~(\ref{dqpertlargeeps}) with the result of Ref.~\onlinecite{Glazman90},
we find that charge smearing by tunneling via the impurity level is
equivalent to that caused by tunneling through an effective barrier with
the conductance $G=(e^2/\pi\hbar) 4\Gamma_g\Gamma_l/\epsilon^2$.
Naturally, this is exactly the conductance through the impurity in the
limit of large $\epsilon$.

As in the case of a simple tunneling barrier\cite{Glazman90,matveev91},
the perturbative result diverges if one of the charging energies, $U_1$ or
$U_{-1}$, approaches zero.  The exact form of the charge smearing around
the degeneracy points can be studied by mapping the system onto a 2-channel
Kondo problem \cite{matveev91}.

Equation (\ref{dqmostgen}) is particularly interesting in the regime where
the impurity level is near resonance, $\epsilon\sim\Gamma$.  Assuming that
the gate voltage is sufficiently far from the degeneracy points, i.e.,
$\epsilon,\Gamma_{l,g}\ll U_{\pm 1}$, we can write Eq.\ 
(\ref{dqmostgen}) in the following compact form,
\begin{equation}
\langle Q\rangle
=e\frac{\Gamma_g}{\pi}\left\{\frac{n(\epsilon)}{U_1}
 -\frac{2-n(\epsilon)}{U_{-1}}\right\} ,
\label{dqusingne}
\end{equation}
with the occupation of the impurity
$n(\epsilon)=1-(2/\pi)\arctan(\epsilon/\Gamma_l)$.  Since $U_1$ and
$U_{-1}$ are of the order of $E_C$, it is clear from this equation that
for a narrow resonance the charge smearing is very small, of the order
$\Gamma_g/E_C\ll 1$.  This is the case even if the impurity level is on
resonance, $\epsilon=0$ and $\Gamma_l=\Gamma_g$, when the transmission at
the Fermi energy is perfect, $T(E_F)=1$. Physically, this result can be
understood in the following way: in order to effectively smear out
charging effects, the grain has to be perfectly coupled to the lead over
an energy interval $\Delta E\gg E_C$ around the Fermi energy. However, in
the case of resonant coupling, the transmission probability is strongly
energy dependent.  A resonant impurity level of width $\Gamma\ll E_C$
leads to perfect transmission only in the very narrow interval $\Delta
E\sim\Gamma$. The transmission at all other energies essentially vanishes.
To our knowledge, all previous work, that predicted charging effects to
completely disappear as soon as there is at least one perfectly
transmitting channel coupling the grain to a lead, assumed that the
coupling is energy independent on the
\begin{figure}[t]
\epsfxsize=.8\columnwidth
\hfill\epsffile[90 84 705 528]{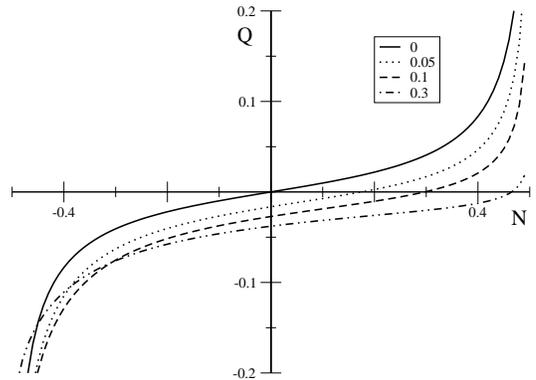}\hfill
\vspace{4ex}
\caption{The smearing of the Coulomb staircase, Eq.\ (\ref{dqmostgen}),
  for different values of $\epsilon/E_C$.
  The coupling strengths have been chosen to
  be $\Gamma_l=\Gamma_g=0.1E_C$.  Drawn is the charge $\langle Q\rangle$ on
  the grain in
  units of the electron charge $e$ as a function of the dimensionless
  gate voltage $N$. The divergences at the degeneracy points $N=\pm 0.5$
  indicate the breakdown of the perturbative result at these points.
  Note the good quantization of the charge of the grain for $-0.4<N<0.4$,
  even if the impurity is on resonance (solid line).\label{stairfig}}
\end{figure}
\noindent
scale of
$E_C$\cite{matveev95,nazarov99}. Our result clearly shows that a possible
energy dependence in the coupling of the grain to the lead can have a
dramatic effect on the shape of the Coulomb staircase.

In Fig.~\ref{stairfig} the smearing of one step of the Coulomb staircase
is shown for different values of the impurity energy $\epsilon$. It is
clearly visible, that even at $\epsilon=0$ the sharp Coulomb blockade step
is only smeared out very little due to the narrow resonance, $\Gamma\ll
E_C$. At $N=0$ the slope of the Coulomb staircase with $\epsilon=0$ is the
same as the one found for a grain coupled through a tunneling barrier of
effective conductance $G=(e^2/\hbar)2\Gamma_g/E_C$ to the lead.
Therefore, away from the degeneracy points, a narrow resonant impurity level
acts similarly to a poorly conducting tunneling barrier as far as charge
smearing is concerned. However, if we approach a degeneracy point,
e.g., $U_1\ll(\Gamma_l+\Gamma_g)$, the smearing due to a resonant level is
very different from the case of a tunneling barrier. At small $U_1$, the
coupling of the grain to the lead in fact is strong and thus the exact shape
of the step of the staircase will differ considerably from the one found in
\cite{matveev91}. Finding the exact shape of the step is a difficult task,
which lies beyond the scope of this paper.

\begin{figure}[t]
\epsfxsize7cm
\hfill\epsffile[90 84 710 533]{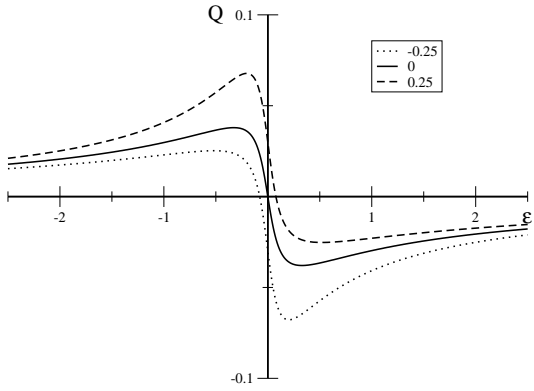}\hfill
\vspace{4ex}
\caption{The charge on the grain in units of $e$  as a
  function of $\epsilon/E_C$ for different values of the gate voltage $N$.
  The couplings are $\Gamma_l=\Gamma_g=0.1E_C$.}
\label{stepfig}
\end{figure}

As the
energy of the impurity is increased, the step of the staircase is pushed
downwards, because the virtual processes of electron tunneling from the
grain onto the impurity become more likely than processes where an
electron from the partially occupied impurity tunnels onto the grain.
However, in the limit of very large $\epsilon$, hopping on the impurity
becomes energetically more and more costly, so that the step moves back up
again.  This is illustrated in Fig.\ \ref{stepfig},\hfill where the charge on
the grain is
\noindent
drawn for different fixed values of the gate voltage $N$ as a
function of the impurity energy $\epsilon$. As the energy of the impurity
crosses zero, the average charge makes an abrupt change from a positive to
a negative value. The width of this jump is $\Delta\epsilon\sim\Gamma_l$.
As can be seen from Eq.\ (\ref{dqusingne}), measuring the average charge
on the grain as a function of the impurity energy $\epsilon$ in this
transition region is a direct measurement of the occupation $n(\epsilon)$.

Up to now we neglected the interaction of the electrons on the impurity.
However, if in an experiment the impurity level is replaced by a second
quantum dot in a heterostructure as suggested in the introduction,
Fig.~\ref{schema}, it has to be much smaller than the large quantum dot,
so that its charging energy $U$ greatly exceeds the charging energy $E_C$
of the dot.  We will show now that a very similar result as
Eq.~(\ref{dqusingne}) is also obtained if we include the strong Hubbard
repulsion of the electrons on the impurity in our Hamiltonian. We add the
following term to the Hamiltonian,
\begin{equation}
H_U=U\hat{n}_\uparrow\hat{n}_\downarrow , \label{intham}
\end{equation}
with $U$ being a very large energy, $U\gg E_C$, and
$\hat{n}_\sigma=a^\dagger_\sigma a_\sigma$.  The strong on-site Coulomb
repulsion will now prohibit double occupation of the impurity level.
To find the average charge on the grain, we can proceed with
the so modified Hamiltonian and rederive Eq.\
(\ref{dqsecondord}). Now the impurity Green's functions in Eq.
(\ref{dqsecondord}) are non-trivial, because the Hamiltonian includes the
on-site interaction (\ref{intham}). We consider a regime where the Coulomb
energies $U_{1}$ and $U_{-1}$ are much larger then $\epsilon$. The Green's
function of the impurity, $\langle a_\sigma^\dagger(t) a_{\sigma}(0)\rangle$,
varies on a time scale $t\sim 1/\epsilon$, whereas the Green's functions
$\langle a_{p\sigma}^\dagger(t) a_{p\sigma}(0)\rangle$ vary on the much
shorter time scale $t\sim 1/U_{-1}$. We can therefore assume that the Green's
function of the impurity is roughly constant in the relevant range of
integration over $t_1$ and $t_2$, and is given by
$n(\epsilon)=\sum_\sigma\langle a_\sigma^\dagger a_\sigma\rangle$. The
integrals in Eq.\ (\ref{dqsecondord}) can then be carried out, and we arrive
at the same result as Eq.\ (\ref{dqusingne}), except that in the presence of
interactions on the impurity $n(\epsilon)$ is the occupation of the impurity
in the Anderson model $H_0+H_{li}+H_U$. As can be seen from Eq.\
(\ref{dqusingne}) and Fig.~\ref{stepfig}, the measurement of the charge on the
grain can be used to determine the occupation $n(\epsilon)$ of the Anderson
impurity.

In this paper we have studied the influence on the Coulomb blockade of a
strong energy dependence of the coupling of a grain to its environment.
The example we investigated was a grain, which was coupled to a lead via a
resonant impurity level. Charge smearing by this coupling has two origins:
first, the sole presence of a nearby impurity can already smear the charge
on the grain even without any coupling to the lead. The second reason for
charge smearing is due to transfer of electrons from the grain over the
impurity to the lead. We showed, that a narrow resonance, $\Gamma\ll E_C$,
is not sufficient to effectively smear out charging effects.  It is worth
noting that our result is not related to the phenomenon of mesoscopic
charge quantization\cite{aleiner98}, which results in small Coulomb
blockade oscillations in a perfectly coupled small quantum dot.  Unlike
Ref.~\cite{aleiner98}, our Coulomb blockade oscillations can be large, and
they do not disappear in the limit of vanishing level spacing in the dot.
We also showed that the charge on the grain can be used to measure the
occupation of the impurity, see Eq.~(\ref{dqusingne}). The easiest way
to experimentally verify our prediction is probably to use a double dot
system in a semiconductor heterostructure, where one of the dots plays the
role of the impurity, Fig.~\ref{schema}.

The authors are grateful to D. Esteve for a stimulating discussion.  T.G.
acknowledges support from the Swiss National Science Foundation. K.M.
acknowledges support by A.P.  Sloan Foundation and NSF Grant No.
DMR-9974435.

\end{multicols}
\end{document}